\documentclass[aps, prl, superscriptaddress, reprint]{revtex4-1}
\usepackage{amsfonts}
\usepackage{graphicx}
\usepackage{amsmath}
\usepackage{times}
\usepackage{amssymb}
\usepackage{color}
\usepackage{bm}
\usepackage{bbm}
\usepackage{array}
\usepackage{natbib}
\usepackage{comment}

\newcommand{\hide}[1]{}
\newcommand{\be}{\begin{equation}}
\newcommand{\bee}{\begin{equation*}}
\newcommand{\ee}{\end{equation}}
\newcommand{\eee}{\end{equation*}}
\newcommand{\bearre}{\begin{eqnarray*}}
\newcommand{\eearre}{\end{eqnarray*}}
\newcommand{\bearr}{\begin{eqnarray}}
\newcommand{\eearr}{\end{eqnarray}}

\begin{document}
\title{Bell correlations at Ising quantum critical points}
\author{Angelo Piga}
\email{angelo.piga@icfo.eu}
\affiliation{ICFO-Institut de Ciencies Fotoniques, The Barcelona Institute of Science and Technology, Av. Carl Friedrich Gauss 3, 08860 Barcelona, Spain}
\author{Albert Aloy}
\affiliation{ICFO-Institut de Ciencies Fotoniques, The Barcelona Institute of Science and Technology, Av. Carl Friedrich Gauss 3, 08860 Barcelona, Spain}
\author{Maciej Lewenstein}
\affiliation{ICFO-Institut de Ciencies Fotoniques, The Barcelona Institute of Science and Technology, Av. Carl Friedrich Gauss 3, 08860 Barcelona, Spain}
\affiliation{ICREA-Instituci\'o Catalana de Recerca i Estudis Avan\c cats, Lluis Companys 23, 08010 Barcelona, Spain}
\author{Ir\'en\'ee Fr\'erot}
\email{irenee.frerot@icfo.eu}
\affiliation{ICFO-Institut de Ciencies Fotoniques, The Barcelona Institute of Science and Technology, Av. Carl Friedrich Gauss 3, 08860 Barcelona, Spain}

\date{\today}

\begin{abstract}
When a collection of distant observers share an entangled quantum state, the statistical correlations among their measurements may violate a many-body Bell inequality, demonstrating a nonlocal behavior. Focusing on the Ising model in a transverse-field with power-law ($1/r^\alpha$) ferromagnetic interactions, we show that a permutationally invariant Bell inequality based on two-body correlations is violated in the vicinity of the quantum-critical point. This observation, obtained via analytical spin-wave calculations and numerical density-matrix renormalization group computations, is traced back to the squeezing of collective-spin fluctuations generated by quantum-critical correlations. We observe a maximal violation for infinite-range interactions ($\alpha=0$), namely when interactions and correlations are themselves permutationally invariant.
\end{abstract}
\maketitle
\textit{Introduction.}
Nonlocal correlations, witnessed by the violation of Bell inequalities (BIs), mark the strongest departure from classical physics that correlated quantum systems may exhibit \cite{brunner_bell_2014}. To violate a BI, entanglement among the individual degrees of freedom is necessary (albeit not sufficient \cite{werner_quantum_1989}). Such quantum correlations are typically fragile against thermalization, especially when considering many degrees of freedom. Nevertheless, thermalization is not always detrimental to entanglement: indeed, quantum critical points (QCPs) \cite{sachdev_quantum_2011} represent a special instance of equilibrium states, where multipartite entanglement is stabilized at all length scales \cite{hauke_measuring_2016}, a feature intrinsically robust at finite temperature in the quantum-critical regime \cite{hauke_measuring_2016, gabbrielli_multipartite_2018, frerot_reconstructing_2019}. In addition to entanglement, are there QCPs which stabilize also nonlocal correlations among the individual components of the system? An important result from quantum information theory shows that all non-product pure states, including those at QCPs, possess bipartite nonlocal correlations \cite{gisin_bells_1991}. Demonstrating and quantifying the presence of nonlocal correlations among a macroscopic number of degrees of freedom is, in general, a very challenging task \cite{brunner_bell_2014}. Nonetheless, a permutationally invariant Bell inequality (PIBI) involving only first and second moments of collective observables was derived recently \cite{tura_detecting_2014, tura_nonlocality_2015}, which is especially revelant for a collection of $N$ qubits. The preparation, in a Bose-Einstein condensate (BEC), of massively entangled states of two-level atoms violating this inequality, was subsequently reported \cite{schmied_bell_2016}.

In the BEC experiment \cite{schmied_bell_2016}, violatation of the PIBI was achieved through the dynamical generation of spin-squeezed states \cite{schmied_bell_2016, pezze_quantum_2018}. On the other hand, spin squeezing is known to be present at the QCP of the transverse-field ferromagnetic (FM) Ising model (TFIM), at least for a sufficiently large number $d$ of spatial dimensions \cite{frerot_quantum_2018, gabbrielli_multipartite_2018}. Here, we investigate nonlocal correlations at the QCP of the TFIM with power-law decaying ($1 / r^\alpha$) interactions, interpolating between infinite-range ($\alpha=0$) and nearest-neighbour ($\alpha \to \infty$) interactions. Besides its fundamental interest as a paradigmatic model for quantum phase transitions, this model has been implemented in various quantum simulators \cite{bernien_probing_2017, zhang2017observation, changetal2018}. We first establish spin squeezing as a necessary condition to violate the PIBI derived in ref.~\cite{tura_detecting_2014, tura_nonlocality_2015}, in a Bell scenario involving identical measurement settings on all qubits. Based on numerical density-matrix renormalization group (DMRG) and analytical linear spin-wave (LSW) computations, we show that spin squeezing is a generic feature close to the QCP, leading to a maximal violation for $\alpha < d$ in the thermodynamic limit. Interestingly, the violation of the PIBI is maximal for all-to-all interactions, where the semi-classical spin-wave theory is exact. Bipartite entanglement entropy (EE), on the other hand, shows the opposite behavior, being maximal for nearest-neighbour interactions.

\textit{Bell inequality violation and spin squeezing.}
We consider a $N$-qubits quantum state, and a Bell scenario in which every qubit can be projectively measured along two possible directions ${\bf n}$ and ${\bf m}$. We aim at certifying the nonlocal nature of the resulting correlations, relying on 1- and 2-body expectation values. More specifically, we consider BIs involving symmetric combinations of such correlators, namely $S_{\bf u} = \sum_{i=1}^N \langle \sigma^{\bf u}_i \rangle$ and $S_{{\bf u}{\bf v}} = \sum_{i\neq j} \langle \sigma^{\bf u}_i \sigma^{\bf v}_j \rangle$ with ${\bf u}, {\bf v} \in \{{\bf n}, {\bf m} \}$, where $\sigma^{\bf u}_i = {\bf u} \cdot \vec{\sigma_i}$, and $\vec{\sigma_i}$ the vector of Pauli matrices. In \cite{tura_detecting_2014}, the following BI was established
\begin{equation}
	W = 1 - \frac{1}{N}S_{\bf n} + \frac{1}{4N}(S_{{\bf n}{\bf n}} - 2S_{{\bf n}{\bf m}} + S_{{\bf m}{\bf m}}) \ge 0 ~,
	\label{ineq6}
\end{equation}
which must be fulfilled by any statistical model obeying Bell locality hypothesis. Given a quantum state, we look for optimal measurement directions $({\bf n}, {\bf m})$ in order to maximally violate inequality \eqref{ineq6}. This optimization can be performed analytically. First, introducing ${\bf a} = ({\bf n} - {\bf m}) / |{\bf n} - {\bf m}|$, defining the collective spin ${\vec J} = \sum_{i=1}^N {\vec \sigma}_i / 2$, and using elementary spin algebra, Eq.~\eqref{ineq6} can be recast in the equivalent form \cite{schmied_bell_2016}
\begin{equation}
	W = 1-|C_{\bf n}| + ({\bf a} \cdot {\bf n})^2 (\zeta_{\bf a}^2 - 1) \ge 0 ~,
	\label{eq_Basel_witness}
\end{equation}
where $C_{\bf n} = \langle J^{\bf n} \rangle / (N / 2) \equiv 1 - r < 1$ and $\zeta_{\bf a}^2 = \langle (J^{\bf a})^2 \rangle / (N / 4)$ are the first and second moments of the collective spin along, respectively, directions ${\bf n}$ and ${\bf a}$, scaled to the coherent spin state values.

\begin{figure}
	\includegraphics[width=1\linewidth]{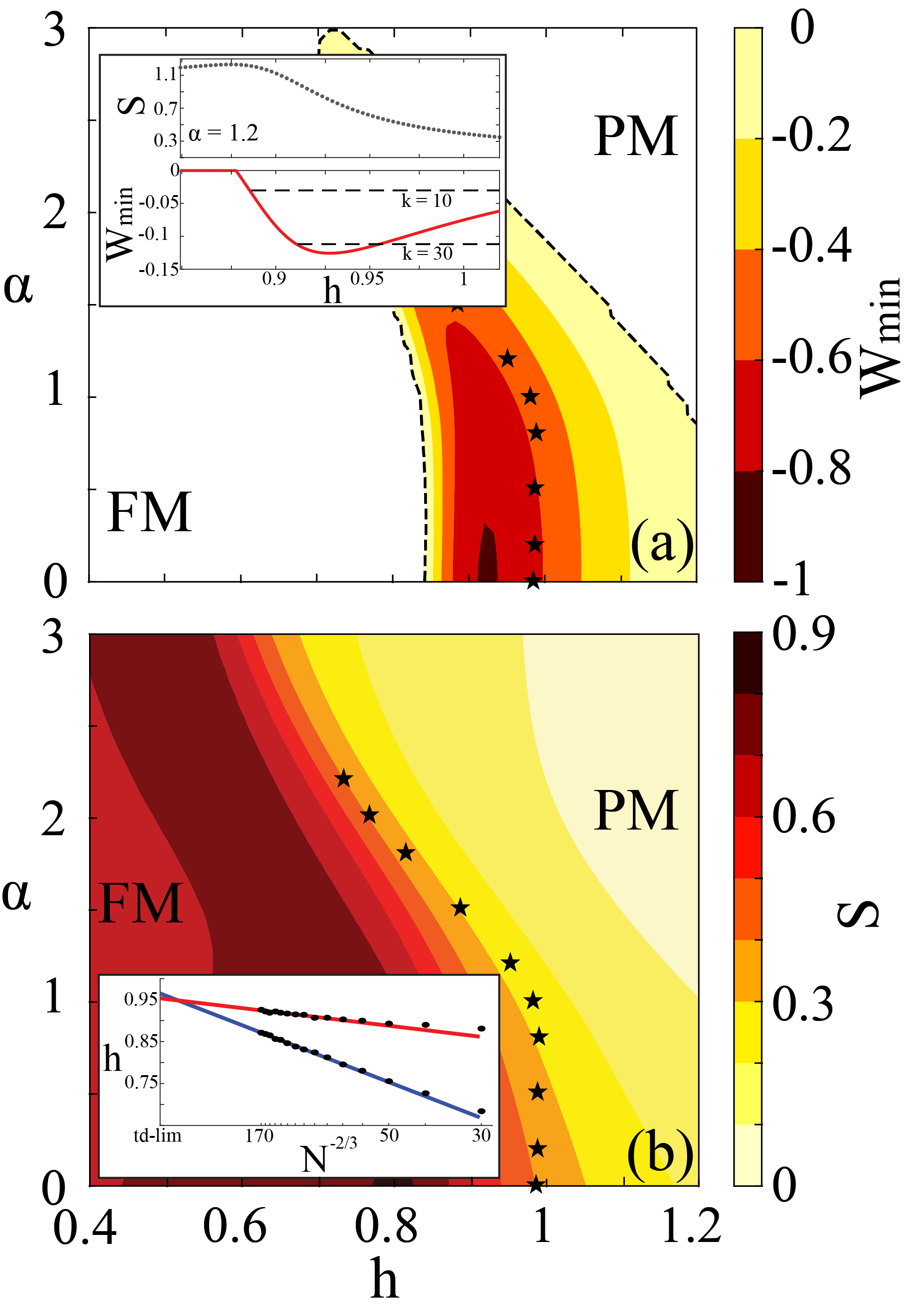}
\caption{(a) BI violation and (b) bipartite EE (b) for the one-dimensional long-range TFIM ($N = 40$). Stars are extrapolations for $N \to \infty$ of the maximum of EE. Inset in (a): bipartite EE and maximal violation of BI Eq.~\eqref{ineq6} for $N = 170$ and $\alpha=1.2$. {Black dashed lines: $k$-producibility bounds ~\cite{aloyetal2019} (see text) for $k=10$ and $k=30$.} Inset in (b): critical point extrapolated for $N\to \infty$. Both maximum of EE and maximal violation of the BI occur for the same transverse-field. Fits of the form $h_c(N) = h_c(\infty) + a N^{-2/3}$ \cite{footnote_caption1}.}
\label{fig_phasediag_1d}
\end{figure}
Notice that measuring the collective spin projectively along ${\bf n}$ and ${\bf a}$ does not realize a Bell scenario, but witnesses the ability to prepare many-spin states exhibiting nonlocality if the spins are individually measured along ${\bf n}$ and ${\bf m}$ \cite{schmied_bell_2016}. Indeed, Eqs.~\eqref{ineq6} and \eqref{eq_Basel_witness} have a different status. On the one hand, Eq.~\eqref{ineq6} allows for a device-independent test of nonlocality, valid even if individual measurement axes are not well-controlled, and even if the individual systems are actually not qubits but have an arbitrary physical structure. The only assumption leading to Eq.~\eqref{ineq6} -- beyond Bell locality hypothesis -- is that two possible measurement settings can be freely chosen on each party, each of which yielding two possible outcomes \cite{tura_detecting_2014, tura_nonlocality_2015}. On the other hand, Eq.~\eqref{eq_Basel_witness} relies on extra physical assumptions: applicability of quantum-mechanical spin algebra and correct calibration of measurement axes \cite{schmied_bell_2016}. 

We define ${\bf z}$ as the mean spin direction: $\langle \vec J \rangle \propto {\bf z}$. If ${\bf a} \cdot {\bf z} \neq 0$, then $\zeta_{\bf a}^2 \propto N$, precluding violation of Eq.~\eqref{eq_Basel_witness} for large $N$. Hence, axis ${\bf a}$ must be chosen perpendicular to ${\bf z}$. Then, the minimal value of $W$ is obtained if ${\bf a}$ is along the direction of minimal variance of ${\vec J}$. Violation of inequality \eqref{eq_Basel_witness} then requires $\zeta_{\bf a}^2 < 1$ (namely, spin squeezing \cite{kitagawa_squeezed_1993, pezze_quantum_2018}), while maintaining the largest possible spin length ($r \ll 1$). Then, we choose ${\bf n} = {\bf z}\cos \phi +  {\bf a}\sin \phi$, yielding $W = (1-\zeta_{\bf a}^2)\cos^2 \phi - (1 - r)\cos \phi + \zeta_{\bf a}^2$. The minimal $W$ is
\begin{equation}
	W_{\min} =  \zeta_{\bf a}^2 - \frac{(1 - r)^2}{4(1 -  \zeta_{\bf a}^2)} > -\frac{1}{4} ~,
	\label{eq_Wmin}
\end{equation}
achieved for $\cos \phi = (1 - r) / [2(1 -  \zeta_{\bf a}^2)]$. The second measurement direction is ${\bf m} = {\bf z}\cos \phi -  {\bf a}\sin \phi$. Maximal violation of inequality \eqref{eq_Basel_witness} is achieved for perfect squeezed states ($\zeta_{\bf a}^2  \to 0$ and $r \to 0$), possible only for $N \to \infty$.

\textit{Ferromagnetic Ising model.} We investigate violation of Eqs.~\eqref{ineq6} and \eqref{eq_Basel_witness} at the QCP of the TFIM, with power-law FM interactions:
\begin{equation}
	{\cal H} = - \frac{1}{\gamma_{\bf 0}} \sum_{i \neq j} \gamma_{ij} S_i^{\bf x} S_j^{\bf x} - h\sum_i S_i^{\bf z}
\end{equation}
where $\gamma_{ij} = |{\bf l}_i - {\bf l}_j|^{-\alpha}$, and $S_i^{{\bf a}={\bf x},{\bf y},{\bf z}} = \sigma_i^{\bf a}/2$ are $s=1/2$ spin operators. $i$ and $j$ run over the sites of a $d-$dimensional square lattice of size $N=L\times (L/2)^{d-1}$ , and ${\bf l}_i$ denotes the position of spin $i$. We introduced $\gamma_{\bf k} = N^{-1}\sum_{i \neq j} \exp[-{\bf k}\cdot ({\bf l}_j - {\bf l}_i)] \gamma_{ij}$, and, to have a well-defined thermodynamic limit also for $\alpha < d$, we normalized the interaction term to $\gamma_{{\bf k} = {\bf 0}}$. Mean-field theory predicts a QCP for $h=h_c=1$, separating paramagnetic (PM) (for $h>h_c$) from FM phases ($h<h_c$). The exact QCP is in general at $1/2 \le h_c \le 1$; in the $d=1$ nearest-neigbour limit, $h_c=1/2$ \cite{sachdev_quantum_2011}.  In the PM phase, spins are aligned along ${ {\bf z}}$; in the FM phase, they sponteneously align along ${\tilde { {\bf z}}} = { {\bf z}}\cos \theta \pm  { {\bf x}}\sin \theta$, with $\cos \theta = h$ in mean-field theory.
At the QCP, fluctuations of the magnetization along ${\bf x}$ diverge as a power-law with the system size, namely $\langle (J^{\bf x})^2 \rangle / N \sim N^{\theta(\alpha)}$ with a critical exponent $\theta(\alpha)$. On the other hand, due to the presence of the transverse-field, the system maintains a finite magnetization along ${\bf z}$, so that $\langle J^{\bf z} \rangle / N = O(1)$. In virtue of Heisenberg inequality for the collective spin, this opens the possibility for squeezing the fluctuations of $J^{\bf y}$, as $\zeta_{\bf y} = \langle (J^{\bf y})^2 \rangle / N \ge O(N^{-\theta(\alpha)})$. While quantum-critical spin squeezing is indeed present when $\alpha = 0$ \cite{DusuelV2004, frerot_quantum_2018}, for nearest-neighbour interactions it is present for $d\ge2$ but absent in $d=1$ \cite{Liuetal2013, frerot_quantum_2018}. FM power-law interactions increase the connectivity of the Ising model, and can be viewed as effectively increasing the physical dimension of the system. Hence, we may expect spin squeezing, as well as the resulting violation of inequality \eqref{ineq6}, to exhibit a non-trivial behavior when varying the power-law exponent $\alpha$ at the QCP. In particular, in $d=1$, we may expect a violation for small values of $\alpha$, but not in the nearest-neighbour limit $\alpha \to \infty$. This scenario is indeed confirmed by our numerical DMRG results \footnote{
Our DMRG algorithm for long-range interactions follows ~\cite{crosswhite2008, frowis2010}.
}, consistently with LSW analytical predictions. 


\begin{figure}
	\includegraphics[width=1\linewidth]{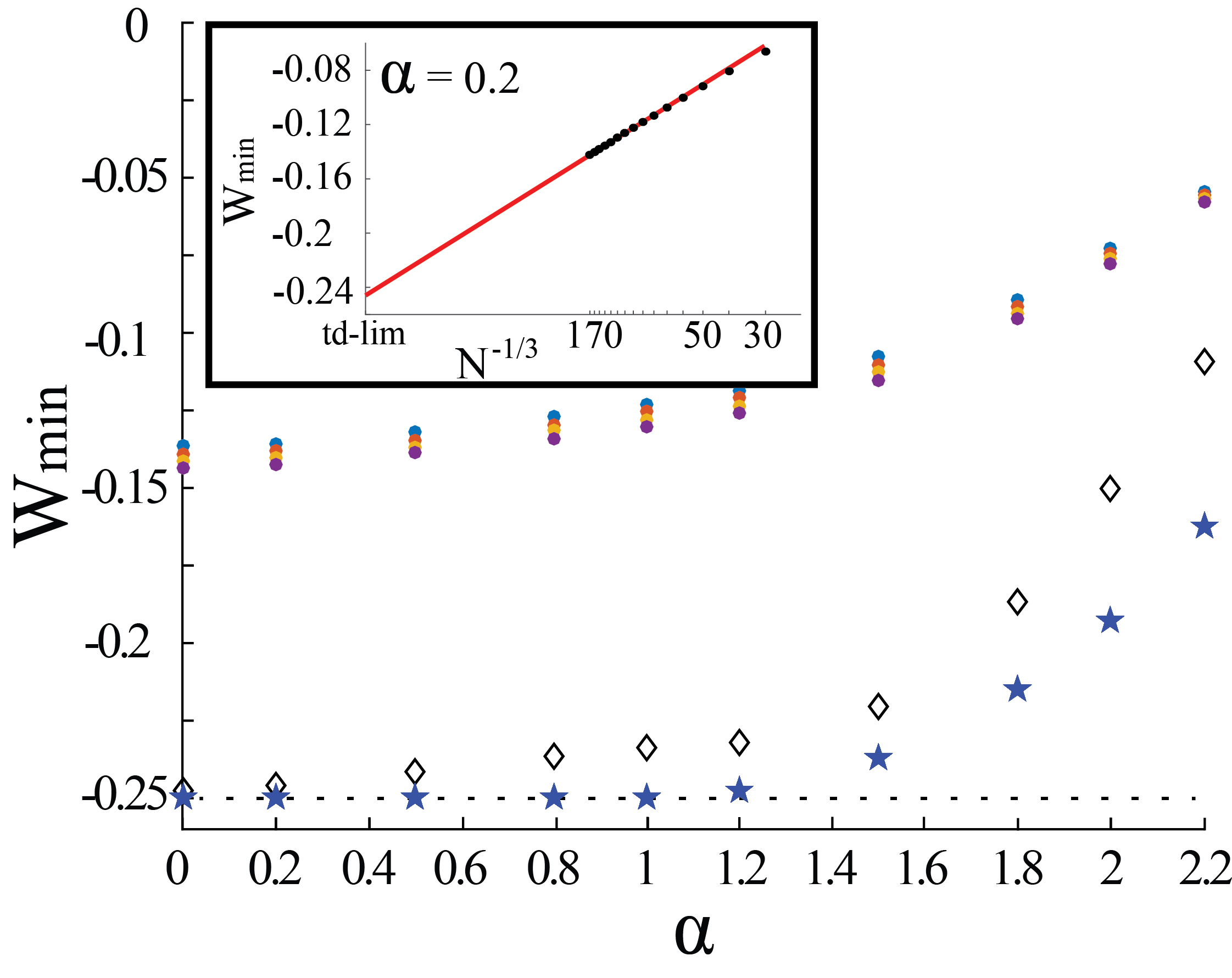}
\caption{BI violation at the QCP of the $d=1$ TFIM, for different values of $\alpha$. Dots: finite size DMRG calculations ($N = 150, 160, 170$). Diamonds: extrapolations for $N \to \infty$ (using $N = 30, 40,\dots,170$). Stars: LSW results ($N = 10^5$).  Inset: extrapolation for $\alpha = 0.2$, of the form $W_{\min}(N) = W_{\min}(\infty) + aN^{-1/3}$ \cite{footnote_caption2}.
}
\label{fig_Wmin_QCP_1d}
\end{figure}

\textit{DMRG results in one dimension.} On Fig.~\ref{fig_phasediag_1d}(a), we plot the maximal violation of the BI [Eq.~\eqref{ineq6}], as a function of the transverse-field $h$ and of $\alpha$. For values of $\alpha\lesssim 3$, nonlocal correlations are detected in the vicinity of the QCP, with maximal violation for $\alpha \to 0$. For $\alpha \gtrsim 3$, no violation is detected, consistently with the quasi-absence of spin squeezing at the nearest-neighbour QCP \cite{Liuetal2013, frerot_quantum_2018}. Fig.~\ref{fig_phasediag_1d}(b) shows von Neumann half-chain EE. Regardless of $\alpha$, for $N\to \infty$, EE is maximal at the QCP. The quantum-critical origin of the BI violation is demonstrated on Fig.~\ref{fig_phasediag_1d}(b), as maximal EE and maximal violation of Eq.~\eqref{ineq6} occur for the same transverse-field in the thermodynamic limit.
On Fig.~\ref{fig_Wmin_QCP_1d}, we plot, varying power-law exponent $\alpha$ and system size $N$, the maximal violation of Eq.~\eqref{ineq6} obtained at the finite-size precursor $h_c(\alpha, N)$ of the QCP (defined as the value of $h$ for which $W_{\min}$ is minimal). For $\alpha < d$ and $N \to \infty$, LSW theory (detailed below) predicts that $W_{\min} \to -1/4$. Due to strong finite-size effects, our extrapolation for $N \le 170$ does not exactly match this prediction (see inset of Fig.~\ref{fig_Wmin_QCP_1d}). However, increasing $\alpha$, we clearly see a weakening violation of Eq.~\eqref{ineq6}, up to $\alpha \gtrsim 3$ where no violation is detected any more.
{
In \cite{aloyetal2019}, it was proven that $k$-producible states \footnote{{
A state $\rho$ is $k$-producible if it is a mixture of tensor products of states involving at most $k$ spins \cite{Toth2012, Hyllusetal2012}. The entanglement depth is the minimal value of $k$ such that $\rho$ is $k$-producible.}
} cannot violate the PIBI by more than a certain bound $b(k)$. Therefore, observing a violation exceeding $b(k)$ certifies an entanglement depth of at least $k+1$. Such bounds are indicated on Fig.~\ref{fig_phasediag_1d}(a, inset). In particular, a violation $W_{\min} \to -1/4$, as observed at the critical point for $\alpha < 1$, certifies a diverging entanglement depth \cite{aloyetal2019}.
}

\textit{Linear spin-wave theory.} Given that FM power-law interactions harden quantum fluctuations about the mean-field ground state, LSW theory is expected to give accurate predictions, especially for small $\alpha$. In fact, we show that LSW theory even becomes exact in the thermodynamic limit for $\alpha < d$. In the following, we choose FM order along $+{\bf x}$. After Holstein-Primakoff (HP) mapping of spin operators to bosonic modes \footnote{
HP mapping: $S_j^z = (\cos \theta) (1 / 2 - b_j^\dagger b_j^{}) - (\sin \theta) (b_j^{} + b_j^{\dagger}) / 2$ ; $S_j^x = (\sin \theta) (1 / 2 - b_j^\dagger b_j^{}) + (\cos \theta) (b_j^{} + b_j^{\dagger}) / 2$ ; $S_j^y = (b_j^{} - b_j^\dagger) / (2i)$. Expressions are valid up to ordrer $O(b_j^3)$. $b_j$ are bosonic operators which, in Fourier space, read: $b_{\bf k} = N^{-1/2}\sum_j \exp(-i{\bf k}\cdot {\bf r}_j) b_j$.  
}, we obtain the LSW Hamiltonian:
\begin{equation}
 {\cal H}_{\rm LSW} = \frac{\max(1, h)}{2}\sum_{\bf k} ({\hat P}_{\bf k}{\hat P}_{-\bf k} + \omega_{\bf k}^2 {\hat X}_{\bf k}{\hat X}_{-\bf k}) ~,
 \label{eq_H_LSW}
\end{equation}
valid up to second order in HP operators. We introduced $\omega_{\bf k} = \sqrt{1 - \gamma_{\bf k} / (h\gamma_{\bf 0})}$ in the PM phase, and $\omega_{\bf k} = \sqrt{1 - h^ 2\gamma_{\bf k} / \gamma_{\bf 0}}$ in the FM phase. In terms of HP operators $b_{\bf k}^{(\dagger)}$ at wave-vector ${\bf k}$, ${\hat X}_{\bf k}$ and ${\hat P}_{\bf k}$ are defined as ${\hat X}_{\bf k} = \frac{b_{\bf k}^{} + b_{-\bf k}^\dagger}{\sqrt{2}}$ and ${\hat P}_{\bf k} = \frac{b_{-\bf k}^{} - b_{\bf k}^\dagger}{i\sqrt{2}}$, such that $[{\hat X}_{\bf k}, {\hat P}_{{\bf k}'}] = i\delta_{{\bf k}, {\bf k}'}$, and $[{\hat X}_{\bf k}, {\hat X}_{{\bf k}'}] = [{\hat P}_{\bf k}, {\hat P}_{{\bf k}'}] = 0$. The LSW Hamiltonian of Eq.~\eqref{eq_H_LSW} is diagonalized by the Bogoliubov rotation 
$\beta_{\bf k} = {\hat X}_{\bf k} \sqrt{\omega_{\bf k} / 2} + i {\hat P}_{-\bf k} / \sqrt{2\omega_{\bf k}}$, such that ${\cal H}_{\rm LSW} = \max(1, h) \sum_{\bf k} \omega_{\bf k} (\beta_{\bf k}^\dagger \beta_{\bf k}^{} + 1/2)$. Written as in Eq.~\eqref{eq_H_LSW}, the physical meaning of LSW mapping is especially transparent. Indeed, the two quadratures ${\hat P}_{\bf k}$ and ${\hat X}_{\bf k}$ represent collective-spin fluctuations in the two directions transverse to the mean spin orientation, namely (in LSW approximation): ${\hat P}_{\bf k} = J^{\bf y}_{\bf k} / \sqrt{N / 2}$ and ${\hat X}_{\bf k} = J^{\tilde{\bf x}}_{-\bf k} / \sqrt{N / 2}$, with $\tilde{\bf x} = {\bf x}\cos \theta - {\bf z}\sin \theta$ and $J_{\bf k}^{\bf u} = \sum_j \exp(i{\bf k}\cdot {\bf r}_j) ({\bf u}\cdot {\vec S}_j)$. Within LSW theory, their fluctuations are simply harmonic, and sectors corresponding to different wave-vectors ${\bf k}$ decouple from each other. Finally, Eq.~\eqref{eq_H_LSW} allows one to directly read the eigenfrequencies of collective-spin fluctuations, namely $E_{\bf k} = \max(1, h) \omega_{\bf k}$. Approaching the QCP at $h=1$, $\omega_{\bf k=0}$ becomes gapless, implying diverging fluctuations of the $\hat X_{\bf 0}$ quadrature (and, correspondingly, squeezing of the $\hat P_{\bf 0}$ quadrature). In terms of collective-spin degrees of freedom (${\vec J} = \sum_i {\vec S}_i \equiv {\vec J}_{\bf k=0}$), one indeed finds
\begin{equation}
	\langle (J^{\tilde{\bf x}})^2 \rangle = \frac{N}{4\omega_{\bf 0}} ~~;~~\langle (J^{{\bf y}})^2 \rangle = \frac{N\omega_{\bf 0}}{4} ~.
	\label{eq_omega0}
\end{equation}
Divergence of order parameter fluctuations (here, $J^{\bf x} = J^{\tilde {\bf x}} \cos \theta$) is a generic signature of critical phase transitions (quantum or thermal). Squeezing of fluctuations transverse to the order parameter (namely of $J^{\bf y}$), on the other hand, is a genuine signature of quantum criticality without a classical analog \cite{frerot_quantum_2018}. Here, it signals the presence of genuine multipartite entanglement at the QCP \cite{pezze_entanglement_2009, frerot_quantum_2018, gabbrielli_multipartite_2018}, yielding maximal violation of the multipartite BI Eq.~\eqref{ineq6}. LSW theory predicts perfect squeezing of $J^{\bf y}$ fluctuations at the QCP ($\zeta_{\bf y} = \omega_{\bf 0} = 0$), so that from Eq.~\eqref{eq_Wmin}, the minimal value of $W$ is simply
\begin{equation}
	W_{\min} = -\frac{(1 - r)^2}{4} ~~\left[\textnormal{at the QCP}\right] ~.
\end{equation}
At LSW level, BI violation at the QCP has thus a very transparent interpretation, involving solely the reduction of the mean spin length by quantum fluctuations.  

LSW predictions are reliable as long as the mean spin length, $1 - r = (2/N) \langle J^{\tilde {\bf z}} \rangle = 1 - (2/N)\sum_{\bf k}  \langle b_{\bf k} ^\dagger b_{\bf k}^{} \rangle$, is moderately reduced by occupation of HP bosonic modes, namely $r \ll 1$. We find $r=(2N)^{-1} \sum_{\bf k} (1 - \omega_{\bf k})^2 / \omega_{\bf k}^{}$. For $\alpha < d$, $\gamma_{{\bf k} \neq 0} / \gamma_{\bf 0} \to 0$ for $N \to \infty$ \cite{frerot_entanglement_2017}, so that $\omega_{{\bf k} \neq 0} \to 1$, and $r \sim (2N)^{-1} (1 - \omega_{\bf 0})^2 / \omega_{\bf 0}^{}$. In other words, all quantum fluctuations apart from those of the collective spin are effectively frozen out. For any $h\neq 1$, we find $r \to 0$: LSW theory is asymptotically exact at any finite detuning from the QCP. The situation is different for $\alpha > d$. On the one hand, away from the QCP, $\omega_{\bf k}$ is gapped, so that $r$ is always finite. The only possible instance of (infrared) divergence is then at the QCP, where $\omega_{\bf k} \sim k^z$ with dynamical exponent $z = \min[1, (\alpha - d)/2]$ \cite{frerot_entanglement_2017}. The condition for infrared divergence of $r$ is then equivalent to the divergence of $\int dk k^{d - 1} / k^z$ at low $k$, \emph{i.e.} to the condition $z \ge d$, met only for $\alpha \ge 3$ ($z=1$) in $d=1$, where logarithmic divergence occurs. Otherwise, $r$ converges for $N \to \infty$ to a finite value, which must satisfy $r \ll 1$ for LSW theory to be reliable \footnote{
For $d=3$, $r_{\max}\approx 0.045$ for $\alpha=\infty$ at the QCP; for $d=2$, $r \lesssim 0.122$: LSW is always reliable for $d=2, 3$. In $d=1$, $r \approx 0.1$ for $\alpha \approx 2$ at the QCP, but already $r\approx 0.3$ for $\alpha=2.4$, indicating a strong effect of quantum fluctuations for large $\alpha$. We complement LSW by DMRG calculations in $d=1$.
}.

\begin{figure}
	\includegraphics[width=1.\linewidth]{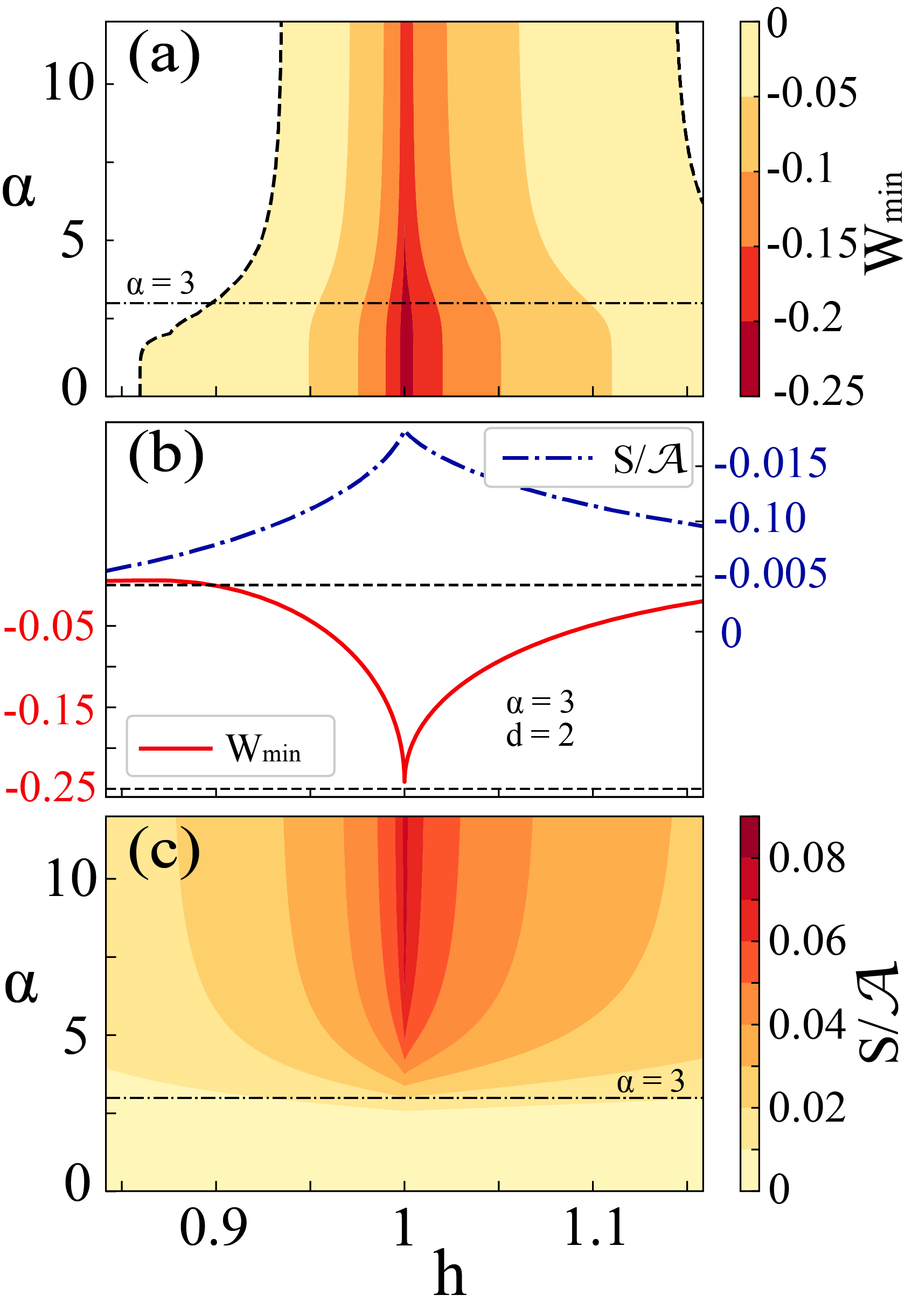}
	\caption{BI violation (a) and bipartite EE (c) for the $d=2$ long-range TFIM. (b) Cut across the line $\alpha=3$ of the phase diagrams [dashed-dotted line on panels (a) and (c)]. Dashed lines: classical ($W_{\min} \ge 0$) and quantum ($W_{\min} \ge -1/4$) bounds of the PIBI. EE is computed for half a torus of size $L_x=200$ times $L_y=100$, and rescaled to the boundary area (${\cal A}=2L_y$).}
	\label{fig_S_phase_diagram_with_inset}
\end{figure}

Remarkably, for $\alpha < d$, $W_{\min} \to -1/4$ in the thermodynamic limit, corresponding to the maximal possible violation of the considered BI \cite{tura_detecting_2014}. This property is illustrated on Fig.~\ref{fig_Wmin_QCP_1d} in $d=1$, and on Fig.~\ref{fig_S_phase_diagram_with_inset}(a) in $d=2$, where $W_{\min}$ is plotted across the phase diagram. It may seem surprising that the limit of infinite-range interactions, leading to a complete suppression of quantum fluctuations at ${\bf k} \neq 0$ in the ground state, is identified as maximally nonlocal. Indeed, in contrast, as shown on Fig.~\ref{fig_S_phase_diagram_with_inset}(c), bipartite EE is strongly suppressed for $\alpha \to 0$, obeying at most a $\log(N)$ scaling for $\alpha < d$  \cite{latorreetal2005} instead of a $L^{d-1}$ (area-law) scaling. This feature should be understood as a specificity of the (permutationally invariant) BI we have considered, rather than an intrinsic property of the many-body state. 
In general, for all $\alpha$, we always find maximal violation of the PIBI at criticality, where bipartite EE is also maximal [Fig.~\ref{fig_S_phase_diagram_with_inset}(b)], demonstrating the quantum-critical origin of the correlations leading to non-locality detection.
Finally, we notice that for $d\ge 2$, in contrast to $d=1$, nonlocal correlations are detected at the QCP for any value of $\alpha$. This observation is consistent with the presence of spin-squeezing for nearest-neighbour interactions in $d\ge 2$ \cite{frerot_quantum_2018}.

\textit{Discussion.} We investigated the violation of a permutationally invariant Bell inequality (PIBI, Eq.~\eqref{ineq6}) induced by a quantum critical point (QCP). We identified spin squeezing -- in a general sense -- as a necessary ingredient to violate the PIBI when identical measurements are performed on a collection of qubits. Focusing on the ground state of the ferromagnetic TFIM, we showed that power-law decaying interactions favor the development of spin squeezing at the QCP, leading to a maximal violation of the PIBI in the limit of infinite-range interactions. Our results are relevant to various experimental platforms implementing the quantum Ising model with power-law interactions, like trapped ions \cite{zhang2017observation}, Rydberg atoms \cite{bernien_probing_2017} and nano-photonic structures \cite{changetal2018}.  In particular, BI violation is expected to be robust against thermal noise \cite{frerot_quantum_2018, gabbrielli_multipartite_2018} and particle losses \cite{tura_detecting_2014}.

Beyond the Ising model considered in this paper, we expect our results to hold for critical points corresponding to the spontaneous breaking an Ising symmetry -- for any range of interactions in $d>1$, and for sufficiently long-range power-law interactions in $d=1$. Extending our study to higher-order symmetries [U(1), SU(2), etc.] is however a non-trivial task, which may require the derivation of novel BIs.

Being invariant under the permutation of any of the $N$ parties involved in the Bell scenario, the BI we have considered is especially suited to investigate nonlocal correlations in $N$-body states themselves permutationally invariant. This absence of spatial structure was indeed realized in the BEC experiment, where $N$ atoms share one spatial mode, as well as in the ground-state of all-to-all interacting models. However, general quantum-critical states, like conventional many-body states, do usually have a non-trivial spatial structure. { As the PIBI only depends on the two-body reduced density-matrix averaged over all pairs, the possibility to capture nonlocal features of QCPs is thus not obvious.} The spatial structure of entanglement, on the other hand, is rather revealed through bipartite Schmidt decomposition, capturing entanglement at a many-body level. Developing further conceptual and technical tools to investigate nonlocal correlations in spatially structured many-body states is an important challenge for ongoing studies \cite{wang_entanglement_2017}.

\begin{acknowledgments}
\textit{Acknowledgements.} We thank T. Roscilde, J. Tura, M. Fadel and E. Tirrito for insightful discussions. We acknowledge the Spanish Ministry MINECO (National Plan
15 Grant: FISICATEAMO No. FIS2016-79508-P, SEVERO OCHOA No. SEV-2015-0522, FPI), European Social Fund, Fundació Cellex, Generalitat de Catalunya (AGAUR Grant No. 2017 SGR 1341 and CERCA Programme), ERC AdG OSYRIS and NOQIA, and the National Science Centre, Poland-Symfonia Grant No. 2016/20/W/ST4/00314. IF acknowledges the Fundació Cellex through a Cellex-ICFO-MPQ postdoctoral fellowship, the Spanish MINECO (QIBEQI FIS2016-80773-P, Severo Ochoa SEV-2015-0522), and the Generalitat de Catalunya (SGR 1381 and CERCA Programme) 
\end{acknowledgments}
\bibliography{biblio_nonlocality_QCP}
\end{document}